 \documentstyle[amstex,amscd,12pt]{article}
\newtheorem{th}{\large\sc Theorem}[section]
\newtheorem{pro}{\large\sc Proposition}[section]
\newtheorem{defi}{\large\sc Definition}[section]
\newtheorem{cor}{\large\sc Corollary}[section]
\begin{document}
\vspace{1cm}
\title{\sc String Equations for the Unitary Matrix Model and the Periodic Flag
Manifold}
\author{
Manuel Ma\~nas\thanks{Research supported by MEC/FLEMING postdoctoral fellowship
GB92 00411668}  $\,$ and Partha Guha\thanks{Research supported by British
Council}\\  The Mathematical Institute,\\
Oxford University,\\ 24-29 St.Giles', Oxford OX1 3LB, \\United Kingdom}
 \maketitle
\begin{abstract}
The  periodic flag manifold  (in the Sato Grassmannian context)  description of
the modified Korteweg--de Vries  hierarchy is used  to
analyse the translational  and scaling self--similar solutions
of this hierarchy. These solutions are characterized by the  string
equations appearing in the double scaling limit of the symmetric unitary matrix
model with boundary terms.
The moduli space  is a double covering of the moduli space in the Sato
Grassmannian for the corresponding self--similar solutions of the Korteweg--de
Vries hierarchy, i.e. of stable 2D quantum gravity.
The potential modified Korteweg--de Vries hierarchy, which can be described in
terms of  a line bundle over the periodic flag manifold, and its self--similar
solutions corresponds  to the symmetric unitary matrix model. Now, the moduli
space is in one--to--one correspondence with a subset of codimension one of the
moduli space in the Sato Grassmannian corresponding to self--similar solutions
of the Korteweg--de Vries hierarchy.
\end{abstract}
\section{Introduction}
In the last few years matrix models have recieved much attention as a
non--per\-tur\-bative formulation of string theory. These models can be
described  in the double scaling limit in terms of solutions to certain
integrable systems. For the Hermitian matrix model (HMM)
 it was found \cite{bk} that in the double scaling limit  the specific heat of
the theory is   a solution to the Korteweg--de Vries (KdV) hierarchy.
This solution must sastisfy also the string equation that it turns to be a
self--similarity condition under Galilean symmetry transformations. This result
was achieved by the use of  orthogonal polynomials on the real line. The string
equation can be written in terms of two scalar differential operators $P,Q$ as
\[
[P,Q]=\text{id}.
\]
 The case of pure gravity  leads to the Painleve I equation. This case
corresponds to the 2--multicritical point of the theory, when one considers the
$k$--multicritical points one finds $2k-2$ order non--linear ODE. Because some
anomalous behaviour of the solutions to the string equation in \cite{djm} it
was proposed an alternative string equation that contains the former one. This
string equation is the self--similarity condition under the local symmetries of
the KdV hierarchy. This theory  is  called 2D stable quantum gravity.

 Later it was shown in \cite{per}, with the use of orthogonal polynomials in
the circle, that for the symmetric unitary matrix model (UMM) in the double
scaling limit the specific heat satisfies  the modified KdV hierarchy and a
string equation. The solutions to this string equation are  scaling
self--similar  solutions to the modified KdV hierarchy.
In the case of  $k$--multicritical point one has a $2k$ ODE which again can be
recasted as
\[
[L,T]=\text{constant}
\]
where $L,T$ are $2\times 2$ matrices of differential operators of order $2k$
and $1$.
In \cite{min} some boundary terms, modelling the presence of quarks,
were added to the model,   the corresponding  string equation  turns out to be
the scaling self--similar condition for the modified KdV hierarchy. This was
connected with 2D stable quantum gravity in \cite{dal3} where the Miura map was
extensively used.

The Sato Grassmannian description \cite{sa} of the solutions was used in
\cite{ks} to characterize the solution to the string equation connected with
the moduli space of complex curves \cite{wi,kon}.  In the papers \cite{sch} a
description of the moduli space for the Galilean self--similar solutions of the
KdV hierarchy in the Sato Grassmannian was given, and in \cite{m} one can find
a more analytical treatment in terms of Stokes parameters. In \cite{gm2} a
complete description, in terms of the initial data for the zero--curvature
1-form, of the moduli space   of self--similar solutions under local symmetries
of the potential KdV hierarchy can be found.
Finally in \cite{abs} one can find a description of the moduli space for the
UMM.

In this paper, following closely the methods of \cite{gm2}, we analyse the
geometrical description of the solutions to the double scaling limit of the UMM
with boundary terms and with out them. It will turn out that the description is
completely different in each case. Our aim is to describe the moduli space of
solutions as a subset of the periodic flag manifold \cite{sw,w} in the Sato
context. We find  that the UMM string equation corresponds to the scaling
self--similarity
condition for the potential modified KdV hierarchy. When border terms are added
the self--similar condition is for the modified KdV hierarchy.
We characterize the moduli space
in terms of the initial data for the corresponding zero--curvature 1--forms
giving in this way a coordinate chart, that happens to be closely connected to
certain algebraic varieties.
The flag manifold is fibered over the Grassmannian and
the moduli space when boundary terms are present is a double covering of the
moduli space for 2D stable gravity. When no boundary terms are present the
moduli space for the UMM is a subspace of codimension one of the former.

Our geometrical description in terms of homogenous spaces and local symmetries
complements that of
\cite{abs}  where  an analysis, based on the fermionic approach, of the moduli
space of the string equation of the UMM with no boundary terms is given.

In the second section we define the modified KdV and potential modified KdV
hierarchies and we give its zero--curvature formulation. We also analyse there
the local symmetries and the  corresponding self--similar conditions, giving a
zero--curvature type formulation of it. In \S 3 we introduce the factorization
problem and the description of these integrable hierarchies in certain
homogeneous spaces. This allows us to study the set of solutions to the string
equations in terms of these homogeneous spaces, essentially a periodic flag
manifold.
In  section four we analyse the moduli space of string equations using the
Sato's periodic flag manifold corresponding to the scaling self--similar
solutions of the modified KdV hierarchy, and a line bundle over this
homogeneous space corresponding to the potential modified KdV hierarchy. In the
final section  we analyse the relation between these moduli spaces of string
equations for the UMM and that of 2D stable quantum gravity.

  \section{Modified KdV hierarchy and string
equations}\setcounter{equation}{0}\setcounter{th}{0}
\setcounter{pro}{0}

We begin this section with the definition of the integrable hierachies known as
the modified KdV (mKdV)  and the potential mKdV hierarchies. They are defined
for
 scalar functions $v,w$  that depends on an infinite number of
variables ${\bf t}:=\{t_{2n+1}\}_{n\geq 0}$,  the local coordinates for
the time manifold $\cal T$. In this convention we adopted $t_1$ to be the space
coordinate, usually denoted by $x$,  and $t_{2n+1}$ with $n>0$ corresponds to
time variables, for example $t_3$ is usally denoted by $t$. For its
construction is very convenient the use of the so called Gel'fand--Dickii
potentials
$R_n[u]$, \cite{gd},
  which are the
coefficients for the expansion of the
kernel of the resolvent  of the associated Schr\"odinger equation with
potential $u$.

\begin{defi}
The modified Korteweg--de Vries hierarchy for $v$ is the following collection
of
compatible equations
\[
\partial_{2n+1}v=\partial_1S_n[v],\;\, n\geq 0
\]
where  $\partial_{2n+1}:=\partial/\partial t_{2n+1}$
and
\[
S_n[v]:=(\partial_1+2v) R_n[u],
\]
where the Gel'fand--Dickii potentials are evaluated  on the Miura
transformation of $v$
\begin{equation}
u=\partial_1v-v^2.\label{miura}
\end{equation}
\end{defi}

 Notice that the potential $u$, given by the Miura map (\ref{miura}),
satisfies the KdV hierarchy
\[
\partial_{2n+1}u=4\partial_1R_{n+1}[u],\;\, n\geq 0.
\]
 The KdV   equation
$4\partial_3u=\partial_1^3u+6u\partial_1u$ follows from the first of its
equations.
  The first  equation of the mKdV hierarchy is the
 mKdV equation
$4\partial_3v=\partial_1^3v-6v^2(\partial_1v)$.

 \begin{defi}
The  potential modified Korteweg--de Vries  hierarchy for the function $w$ is
the following set of equations
\[
\partial_{2n+1}w= S_n[v],\,\; n\geq 0
\]
where
\[
v:=\partial_1 w,
\]
\end{defi}

Observe that if $w$ is a solution to the potential mKdV hierarchy then
$v=\partial_1w$ is a solution to the mKdV hierarchy. The  potential mKdV
equation is  $4\partial_3w=\partial_1^3w-2(\partial_1w)^3$.

These integrable hierarchies are equivalent to zero--curvature conditions,
which turns out to be an essential feuture of its integrability condition.
Novikov \cite{n1} gave for the KdV equation
a zero--curvature representation in terms of a differential 1--form
$\chi(\lambda)$
that depends on a complex spectral parameter $\lambda\in{\Bbb C}$.
The KdV hierarchy has a similar formulation. Let
$\chi$ be  the
 1--form on $\cal T$ defined by
\[
\chi(\lambda):=\sum_{n\geq 0}L_{2n+1}(\lambda)dt_{2n+1},
\]
where
\[
L_{2n+1}(\lambda):=
\left(\begin{array}{cc}
-\frac{1}{2}\partial_1\rho_n(\lambda)&\rho_n(\lambda)\\
(\lambda-u)\rho_n(\lambda)-\frac{1}{2}\partial_1^2\rho_n(\lambda)&
\frac{1}{2}\partial_1\rho_n(\lambda)
\end{array}\right),
\]
with
\begin{equation}
\rho_n(\lambda):=2\sum_{m=0}^n \lambda^mR_{n-m}[u].\label{r}
\end{equation}
Then, the KdV hierarchy is equivalent to the zero--curvature condition,
\[
[d-\chi,d-\chi]=0,
\]
where $d$ is  the exterior derivative
$d:=\sum_{n\geq 0}dt_{2n+1}\partial_{2n+1}$.
For the mKdV hierarchy there is an equivalent statement.
\begin{pro}
The 1--form
\[
\xi:=da\cdot a^{-1}+\mbox{Ad}a\ \chi,
\]
where
\[
a:=\left(\begin{array}{cc}1&0\\v&1\end{array}\right),
\]
has zero--curvature if and only if $v$ is a solution of the mKdV hierarchy.
This 1--form can be represented as
\[
\xi(\lambda):=\sum_{n\geq 0}\ell_{2n+1}(\lambda)dt_{2n+1},
\]
whith
\begin{equation}
\ell_{2n+1}(\lambda):=
\left(\begin{array}{cc}
-(\partial_1+2v)\rho_n(\lambda)/2&\rho_n(\lambda)\\
\lambda \rho_n(\lambda)-  \partial_1(\partial_1+2v)(\rho_n(\lambda)/2-R_n)&
(\partial_1+2v)\rho_n(\lambda)/2
\end{array}\right).\label{ell}
\end{equation}
with $\rho$ given in Eq. {\rm (\ref{r})}
\end{pro}
{\bf Proof:}
It follows from the equation
\[
\partial_{2n+1}\ell_1-\partial_1\ell_{2n+1}+[\ell_1,\ell_{2n+1}]=0.
\]
$\Box$

One can equally proof the following
\begin{pro}
The potential mKdV hierarchy is equivalent to the zero--curvature condition for
the 1--form
\[
\eta=db\cdot b^{-1}+\mbox{Ad}b\ \xi,
\]
where
\[
b:=\left(\begin{array}{cc}\exp(w)&0\\0&\exp(-w)\end{array}\right).
\]

Now
\[
\eta(\lambda):=\sum_{n\geq 0}\tilde\ell_{2n+1}(\lambda)dt_{2n+1},
\]
whith
\begin{equation}
\tilde\ell_{2n+1}(\lambda):=
\left(\begin{array}{cc}
-(\partial_1+2v)(\rho_n(\lambda)/2-R_n)&e^{2w}\rho_n(\lambda)
\\e^{-2w}(\lambda\rho_n(\lambda)-  \partial_1
(\partial_1+2v)(\rho_n(\lambda)/2-R_n)) &
(\partial_1+2v)(\rho_n(\lambda)/2-R_n)
\end{array}\right).\label{tildell}
\end{equation}
\end{pro}

 Let us now consider the symmetries defined by
translations and scaling transformations.

The infinite set of
translational symmetries are the isospectral
symmetries of these hierarchies in the sense that they preserve the
associated spectral problem, i.e. the Schr\"odinger equation for $u$.
In fact the  flows in the hierarchies are defined
by the generators $\partial_{2n+1}$ of  translations.
\begin{defi}
Let be
\[
\vartheta({\bf t}):={\bf t}+\mbox{\boldmath$\theta$},
\]
the action of translations, where
\[
\mbox{\boldmath$\theta$}:=\{\theta_{2n+1}\}_{n\geq 0}
\in{\Bbb C}^{\infty},
\]
are the shifts of the time variables.
\end{defi}

We have a local
action of the abelian group ${\Bbb C}^{\infty}$ over the time manifold
$\cal T$. The following is obvious
\begin{pro}
If $v,w$ are solutions
to the mKdV and potential mKdV hierarchies respectively then so are
$\vartheta^{\ast}v,\vartheta^{\ast}w$.
\end{pro}

For the scaling symmetry  we have
\begin{defi}
The scaling transformation is
\[
\varsigma_\sigma({\bf t}):=
\{e^{\sigma(n+\frac{1}{2})}t_{2n+1}\}_{n\geq 0}
\]
where $\sigma\in{\Bbb C}$.
\end{defi}

We have  an additive local action of $\Bbb C$ over
$\cal T$.
One can easily show that
\begin{pro}
If $v,w$ are
 solutions of the mKdV an potential mKdV hierarchies respectively then so are
$e^{\sigma/2 }\varsigma_\sigma^* v,\varsigma_\sigma^{\ast}w$.
\end{pro}

The related fundamental vector fields, infinitesimal generators of the
action of translation and scaling   transformations are
given by
\[
\partial_{2n+1},n\geq 0,
\,\; \mbox{\boldmath$\varsigma$}
=\sum_{n\geq 0}(n+\frac{1}{2})t_{2n+1}\partial_{2n+1}
\]
respectively. They generate the linear space ${\Bbb C}\{ \partial_{
2n+1}, \mbox{\boldmath$\varsigma$} \}_{n\geq 0 }$ which  is
the Lie algebra of local symmetries of the mKdV and potential mKdV hierarchies,
the Lie brackets are
\[
[\partial_{2n+1},\mbox{\boldmath$\varsigma$}]=(n+\frac{1}{2})\partial_{2n+1}.
\]
We have the very important notion
\begin{defi}
 A self--similar solution under any of the mentioned symmetries is a solution
which remains invariant under the corresponding transformation.
\end{defi}
Consider the following vector field belonging to this
Lie algebra
\begin{equation}
X:=
\mbox{\boldmath$\vartheta$} +\sigma\mbox{\boldmath$\varsigma$}, \label{X}
\end{equation}
with
\[
\mbox{\boldmath$\vartheta$}=\sum_{n\geq 0} \theta_{2n+1}
\partial_{2n+1},
\]
defining  a superposition of translation  and scaling
transformations.
If $v$ is a solution of the mKdV hierarchy then the function
\[
 \exp(\sigma/2+X)v
\]
is a solution as well.
In what follows it will be convenient the
\begin{defi}
Let us denote
\[
{\cal R}:=\sum_{n\geq 0}(n+\frac{1}{2})t_{2n+1}R_n.
\]
\end{defi}
then we have,

\begin{th}
A solution $v$ of the  mKdV hierarchy
is self--similar under the vector field $X$
if and only if
it satisfies the generalized string equation
\begin{equation}
 \partial_1(\partial_1+2v)\left(\sum_{n\geq 0}\theta_{2n+1}R_n+\sigma{\cal
R}\right) =0.\label{sssecu}
\end{equation}
\end{th}
{\bf Proof:}
A solution $v$ of the   mKdV hierarchy is
self--similar under $X$    if
\begin{equation}
(\mbox{\boldmath$\vartheta$}+
\sigma\mbox{\boldmath$\varsigma$}
+\frac{\sigma}{2})v=0.\label{sspp}
\end{equation}
recalling the mKdV hierarchy one can show that this equation is actually
equivalent to (\ref{sssecu}).$\Box$

The theorem above implies
\[
(\partial_1+2v)\left(\sum_{n\geq 0}\theta_{2n+1}R_n+\sigma{\cal
R}\right)=c(t_3,t_5,\dots)+\frac{\sigma}{4},
\]
but
\[
\partial_{2n+1}c=\sum_{n\geq
0}(\theta_{2n+1}+\sigma\,(n+\frac{1}{2})t_{2n+1})\partial_{2m+1}S_n+
\sigma\,(m+\frac{1}{2})S_m.
\]
Using the commuting flow condition $[\partial_{2n+1},\partial_{2m+1}]w=0$ one
realizes that the above equation can be written as
\[
\partial_{2m+1}c=(X+\sigma\,(m+\frac{1}{2}))S_m,
\]
because $v$ is self--similar  the right hand side of this equation vanishes,
hence
$c$ is a constant.

\begin{cor}
The solution $v$ of the  mKdV hierarchy
is self--similar under   $X$
if and only if
 \begin{equation}
 (\partial_1+2v)\left(\sum_{n\geq 0}\theta_{2n+1}R_n+\sigma{\cal R}\right)
=c+\frac{\sigma}{4},\label{ecuc}
\end{equation}
for some complex number $c$.
\end{cor}

 For the potential mKdV hierarchy we have
\begin{th}
A solution $w$ of the potential mKdV hierarchy
is self--similar under the vector field $X$
if and only if
 $v=\partial_1 w$ satisfies Eq. {\rm (\ref{ecuc})} with $c=0$ .
\end{th}

{\bf Proof:}
The self--similar condition is
\[
(\mbox{\boldmath$\vartheta$}+
\sigma\mbox{\boldmath$\varsigma$})w=0
\]
which, using the hierarchy equations, gives the desired result.$\Box$

Notice that  when $\theta(\lambda)=a(N+1/2)\lambda^N$
the translation term in the string equation is removed
if we transform the time  coordinates as follows:
$t_{2n+1}\mapsto t_{2n+1}+a\delta_{nN}$. This allows us to  study the solutions
out from singularities. Observe also that given a self--similar solution $w$ of
the potential mKdV hierarchy then $v=\partial_1 w$ is self--similar with $c=0$.
But given a self--similar solution $v$ of the mKdV hierarchy there is no
self--similar $w$, solution of the potential mKdV hierarchy such that
$v=\partial_1w$, unless $c=0$. The point is that the string equation for the
mKdV hierarchy only implies $\varsigma_\sigma^{\ast}w({\bold t})=w({\bold
t})+w_0(\sigma)$.

The self--similar condition for the potential mKdV hierarchy is the string
equation that appears in \cite{per} for the double scaling limit of the UMM. In
this case $c=0$, but when the self--similar condition is required for the mKdV
hierarchy   there is no need to
 confine $c=0$, this is the case for the double scaling limit of the UMM with
an additional boundary term that models the presence of $c$ flavours of quarks
\cite{min,dal3}.

The general self--similarity condition can be reformulated
 as a zero--curvature type
condition. This approach is closely connected with the isomonodronic technique
employed in \cite{m}. We  define the outer derivative
\begin{equation}
\delta:=\sigma(\lambda\frac{d}{d\lambda}+
\frac{1}{4}\mbox{ad}H),\label{der}
\end{equation}
where
$H=\sigma_3$ is the  diagonal Pauli matrix, observe that $\delta$ is
proportional to the derivation defining the principal grading of the affine Lie
algebra $A^{(1)}_1$,
and
\begin{equation}
M:=\langle\xi,X\rangle,\ \tilde M:=\langle\eta,X\rangle\label{m},
\end{equation}
Here $\langle\cdot,\cdot\rangle$  is the standard pairing between
1--forms and vector fields.
Then one has,
\[
\,
\]
\begin{th}
$\,$
\begin{enumerate}
\item The zero--curvature type condition
\begin{equation}
[d-\xi,\delta-M]=0\label{zcss}
\end{equation}

is equivalent to the generalized
string equation {\rm (\ref{sssecu})}.
\item  The equation
\[
[d-\eta,\delta-\tilde M]=0
\]
is equivalent to Eq. {\rm (\ref{ecuc})} with $c=0$.
\end{enumerate}
\end{th}
{\bf Proof:} For the 1--form $\xi$ this follows from the condition
\[
\exp(t\delta)\xi=\exp(tX)\xi,
\]
that is equivalent to
\[
\delta\xi=L_X\xi,
\]
where $L_X$ denotes the Lie derivative along the vector field $X$.
But
\[
L_X\xi=(i_Xd+di_X)\xi,
\]
and  recalling the zero--curvature condition for $\xi$, we obtain the
desired result.  The same argumentation holds for $\eta$.$\Box$

This theorem is the key for the analysis of the moduli space of the string
equation.

  \section{Homogeneous spaces and  the string equations}
\setcounter{equation}{0} \setcounter{th}{0}\setcounter{pro}{0}
In this section we use the periodic flag manifold  $\mbox{Fl}^{(2)}$
description of the
mKdV flows, \cite{w,ps}, in order to characterize
geometrically the string equations for
the self--similar solutions of the  mKdV
hierarchy. We also analyse the string equation for the potential mKdV
hierarchy,
not in the periodic flag manifold but in some line bundle over
$\text{Fl}^{(2)}$.
These manifolds appears when one considers  certain factorization problems in
loop groups.

Recall that $\xi$ defines a 1--form with values in the loop algebra
$L{\frak sl}(2,{\Bbb C})$ of smooth maps from the circle
$S^1:=\{\lambda\in{\Bbb C}: |\lambda|=1\}$ to
the simple Lie algebra ${\frak sl}(2,{\Bbb C})$, traceless $2\times 2$ complex
matrices.
We define  an infinite set of commuting flows in the corresponding
loop group $LSL(2,{\Bbb C})$
\begin{equation}
\psi({\bf t},\lambda):=\sigma({\bf t},\lambda)\cdot g(\lambda)\label{psi}
\end{equation}
where $g$ is the
initial condition and
\begin{equation}
\sigma({\bf t},\lambda):=\exp(\sum_{n\geq
0}t_{2n+1}\lambda^nJ(\lambda)),\label{sigma}
\end{equation}
with
\begin{equation}
J(\lambda):=\lambda F+E\label{J}
\end{equation}
 in terms of the standard Cartan-Weyl basis
$\{E,H,F\}$ for ${\frak sl}(2,{\Bbb C})$,
i.e.
\[
E=\left(\begin{array}{cc}0&1\\0&0\end{array}\right),\:
H=\left(\begin{array}{cc}1&0\\0&-1\end{array}\right),\;
F=\left(\begin{array}{cc}0&0\\1&0\end{array}\right),
\]
notation that will be used in the rest of the paper.

Now, we introduce some definitions, the notation is that of \cite{ps}. Denote
by $L^+SL(2,{\Bbb C})$ those loops which have a holomorphic
extension to the interior of $S^1$,  by
$L^-SL(2,{\Bbb C})$ those which
extend analitically to the exterior of the circle, and by $L^-_1SL(2,{\Bbb
C})\subset L^-SL(2,{\Bbb C})$ the subset of those extensions  which are
normalized
by the identity at $\infty$. Consider the subgroup
$B^+SL(2,{\Bbb C})$  of loops of $L^+SL(2,{\Bbb C})$ such that its holomorphic
extensions to the interior of $S^1$ when evaluated at the origin belongs to the
standard Borel group of $SL(2,{\Bbb C})$, that is the upper triangular $2\times
2$ matrices with unity determinant.
The group $N^+SL(2,{\Bbb C})$ is defined analogously but now the Borel subgroup
is replaced by the standard nilpotent group, i.e.  upper triangular $2\times 2$
matrices with $1$ in the diagonal. The subgroup $B^-SL(2,{\Bbb C})$ is the set
of loops of $L^-SL(2,{\Bbb C})$ such that its holomorphic extension to the
exterior of $S^1$ when evaluated at infinity  belongs to the set of $2\times 2$
lower triangular matrices with unity determinant, when we ask to the elements
of the diagonal to be equal to 1 we have the subgroup $N^-SL(2,{\Bbb C})$.

The factorization problem
\begin{equation}
\psi=\psi_-^{-1}\cdot\psi_+,\label{fac}
\end{equation}
where $\psi_-\in N^-SL(2,{\Bbb C})$ and $\psi_+\in
B^+SL(2,{\Bbb C})$,
for $\psi({\bf t})$
is  connected with the mKdV hierarchy.
The element $\psi_-$ can be parametrized by a function $v$, in such a way that
$\psi_-$ is a solution to the
factorization problem if and only if $v$ is a solution to the  mKdV hierarchy,
therefore
\begin{equation}
\xi:=d\psi_+\cdot\psi_+^{-1}=
P_+\mbox{Ad}\psi_-\left(\sum_{n\geq
0}\lambda^nJ(\lambda)dt_{2n+1}\right)\label{faco}
\end{equation}
is
the zero--curvature 1--form for the mKdV equation \cite{gm1}.
Here $\mbox{id}=P_++P_-$ is the resolution of the identity related to the
spliting
\[
L{\frak sl}(2,{\Bbb C})=B^+{\frak sl}(2,{\Bbb C})\oplus
N^-{\frak sl}(2,{\Bbb C}).
\]

Similarly, if we consider the factorization problem
\[
\psi=\tilde\psi_-^{-1}\cdot\tilde\psi_+,
\]
  with  $\tilde\psi_-\in B^-SL(2,{\Bbb C})$ and $\tilde\psi_+\in
N^+SL(2,{\Bbb C})$,
for $\psi({\bf t})$
we find the potential mKdV hierarchy.
Now, $\tilde\psi_-$ can be parametrized by a function $w$, such  that
$\tilde\psi_-$ is a solution to the
factorization problem if and only if $w$ is a solution to the  potential mKdV
hierarchy, thus
\[
\eta:=d\tilde\psi_+\cdot\tilde\psi_+^{-1}=
\tilde P_+\mbox{Ad}\tilde\psi_-\left(\sum_{n\geq
0}\lambda^nJ(\lambda)dt_{2n+1}\right)
\]
is
the zero--curvature 1--form for the potential mKdV equation \cite{gm1}.
The resolution  $\mbox{id}=\tilde P_++\tilde P_-$ is associated with the
decomposition
\[
L{\frak sl}(2,{\Bbb C})=N^+{\frak sl}(2,{\Bbb C})\oplus
B^-{\frak sl}(2,{\Bbb C}).
\]

One can conclude from these considerations that
the projection of the commuting flows
$\psi({\bf t})$ on the periodic flag
manifold \cite{ps,w}
\[
LSL(2,{\Bbb C})/
B^+SL(2,{\Bbb C})\cong \mbox{Fl}^{(2)},
\]
can be described   in terms of the mKdV hierarchy.

We must remark that $g$ determines a point in the periodic flag manifold
up to the gauge freedom $g\mapsto g\cdot h$, where $h\in
B^+SL(2,{\Bbb C})$.
A solution of  the mKdV hierarchy
 does  not change when $g(\lambda)\mapsto\exp(\beta(\lambda)
J(\lambda))\cdot g(\lambda)$
if $\exp(\beta J)\in N^-SL(2,{\Bbb C})$.
We can say that the  moduli
space for the KdV hierarchy contains the double coset space
\[
{\cal M}:=\Gamma_-\backslash LSL(2,{\Bbb C})/B^+SL(2,{\Bbb C}) \]
where $\Gamma_-$
is the abelian subgroup with Lie algebra
${\Bbb C}\{\lambda^nJ(\lambda)\}_{n<0}$, \cite{w}.

The potential mKdV hierachy describes the projection of these commuting flows
over
\[
LSL(2,{\Bbb C})/N^+SL(2,{\Bbb C}),
\]
a line bundle over the periodic flag manifold $\text{Fl}^{(2)}$.
Being the moduli space
\[
\tilde{\cal M}:=\Gamma_-\backslash LSL(2,{\Bbb C})/N^+SL(2,{\Bbb C}).
\]

Let us now try to find for which initial conditions $g$
one gets
self--similar solutions, i.e. points in these homogeneous manifolds  that are
connected to self--similar
solutions of the mKdV
hierarchy and to the potential mKdV hierarchy.

Recall that we have the derivation $\delta
\in  \mbox{Der}B^+{\frak sl}(2,{\Bbb C}),\,
\mbox{Der}N^+{\frak sl}(2,{\Bbb C})$ defined in (\ref{der})
and the vectors  $M({\bf t})\in B^+{\frak sl}(2,{\Bbb C}), \tilde M({\bf t})\in
N^+{\frak sl}(2,{\Bbb C})$
defined in (\ref{m}).
If we denote by
\begin{equation}
\theta(\lambda):=\sum_{n\geq 0}\theta_{2n+1}\lambda^n\label{t}
\end{equation}
then it follows
\newpage
  \begin{th}$\,$
\begin{enumerate}
\item If the initial condition $g$ satisfies the
equation
\begin{equation}
\delta g\cdot g^{-1}+{\rm Ad}g K=\theta J,
\label{ecu}
\end{equation}
for  some $K\in B^+{\frak sl}(2,{\Bbb C})$, and $\theta, J$ are given by
{\rm(\ref t), (\ref J)}.
then
the corresponding solution to the  mKdV hierarchy satisfies the
  string equation {\rm (\ref{sssecu})}, i.e. Eq. {\rm (\ref{ecuc})}.
\item  If $g$ satisfies the  Eq. (\ref{ecu})  for some $K\in N^+{\frak
sl}(2,{\Bbb C})$ then the associated solution $w$ to the potential mKdV
hierarchy is self--similar under the vector field $X$ defined in {\rm
(\ref{X})} and so $v=\partial_1 w$ is solution to {\rm (\ref{ecuc})} with
$c=0$.
\end{enumerate}
\end{th}

{\bf Proof:}
We proof the first statement.
For $\xi=d\psi_+\cdot\psi_+^{-1}$  we observe that
the equation (\ref{zcss})
holds if and only if
\begin{equation}
M=\delta\psi_+\cdot\psi_+^{-1}+\mbox{Ad}\psi_+K,\label{mm}
\end{equation}
for some $K\in B^+{\frak sl}(2,{\Bbb C})$. This, together with
the factorization problem (\ref{fac}), implies the relation
\[
M=\delta\psi_-\cdot\psi_-^{-1}+\mbox{Ad}\psi_-(\delta\sigma\cdot\sigma^{-1}+
\mbox{Ad}\sigma(\delta g\cdot g^{-1}+\mbox{Ad}g K)).
\]
Now, $M({\bf t})\in B^+{\frak sl}(2,{\Bbb C})$ and Eq. (\ref{ecu})
gives
\[
M=P_+\mbox{Ad}\psi_-(\delta\sigma\cdot\sigma^{-1}+
\mbox{Ad}\sigma(\theta J)).
\]
 But, as  can be easily shown
\[
M=P_+\mbox{Ad}\psi_-\left(\theta(\lambda)+\sigma\sum_{n\geq 0}
(n+\frac{1}{2})t_{2n+1}\lambda^n\right)J(\lambda).
\]
Taking into account Eq. (\ref{faco}) we recover  (\ref{m}) and therefore the
 string equation is satisfied.
The second statement can be proof as above but replacing $B^+SL(2,{\Bbb C})$ by
$N^+SL(2,{\Bbb C})$. $\Box$

\section{Description of the  moduli space of the string equations}
\setcounter{equation}{0}\setcounter{th}{0}
\setcounter{pro}{0}

Now, we shall give a description of the points in the periodic flag manifold
corresponding to self--similar solutions of the mKdV hierarchy. The periodic
flag manifold $\mbox{Fl}^{(2)}$, \cite{ps,w} is the set of pairs $(V,W)$ of
subspaces in the Hilbert space ${\cal H}=L^2(S^1,{\Bbb C})$ such that they
belong to the Segal--Wilson Grassmannian \cite{sw} and  satisfy the periodicity
condition
$\lambda^2W\subset\lambda V\subset W$.
In the Segal--Wilson framework  only the family of solutions related through
the Miura map to the Adler--Moser rational solutions of the KdV hierarchy
\cite{am} appears as self--similar solutions. A much more large family lies in
the Sato extension of the periodic flag manifold, where    ${\cal H}={\Bbb
C}[[\lambda^{-1},\lambda]$ and the subspaces belongs to the Sato Grassmannian
\cite{sa}. Therefore, we shall consider the Sato periodic flag manifold
$\mbox{Fl}^{(2)}$.
The statements of the previous section which are rigorous in the
Segal--Wilson
case, can be extended to the Sato frame if the formal groups $N^-SL(2,{\Bbb
C}), B^-SL(2,{\Bbb C})$
are considered only
when acting by its adjoint action or by gauge transformations
in the formal Lie algebra
${\frak sl}(2,{\Bbb C})[[\lambda^{-1},\lambda]$.
In this context  Eqs.(\ref{zcss}), (\ref{mm}) and (\ref{ecu})  still holds.

To connect the results of the previous section  with this description we write
\[
g=\left(\begin{array}{cc}\varphi_1&\tilde\varphi_1\\
\varphi_2&\tilde\varphi_2\end{array}\right),
\]
with
$\varphi_1\tilde\varphi_2-\tilde\varphi_1\varphi_2=1$, and
introduce the notation
\[
\Phi=\left(\begin{array}{c} \varphi_1\\ \varphi_2\end{array}\right),\ \
\tilde\Phi=\left(\begin{array}{c} \tilde\varphi_1\\ \tilde\varphi_2
\end{array}\right).
\]
Define also the map \cite{ps,sw} $\Phi\mapsto\varphi:=T\Phi$ where
$(T\Phi)(\lambda):=
\lambda\varphi_1(\lambda^2)+\varphi_2(\lambda^2)$.

Notice that for each equivalence class in $\cal M$ an element $g$
can be taken such that  $\ln g\in {\Bbb C}F\oplus{\frak sl}(2,{\Bbb
C})[[\lambda^{-1})$,
and that any element in the coset $g\cdot B^+SL(2,{\Bbb C})$ gives the same
point in the periodic flag manifold.

Since $\sigma\left|_{{\bf t}=0}\right.=\mbox{id}$, (\ref{sigma}), it follows
from
(\ref{fac}), (\ref{psi}) that $\psi_+\left|_{{\bf t}=0}\right.=\mbox{id}$
and Eq. (\ref{mm}) gives
\[
K=M\left|_{{\bf t}=0}\right..
\]
But, from (\ref{m}) we have
\[
K=\langle\xi\left|_{{\bf t}=0}\right.,\mbox{\boldmath$\vartheta$}\rangle,
\]
where we have taken into account that (\ref{X}) implies
\[
X\left|_{{\bf t}=0}\right.=\mbox{\boldmath$\vartheta$}.
\]
{}From these considerations we conclude
 \begin{th}
The points $(V,W)$ in the Sato periodic
flag manifold ${\rm Fl}^{(2)}$ corresponding to  self--similar solutions of the
mKdV hierarchy are given by
\begin{eqnarray*}
&&V={\Bbb
C}\{\lambda^{-1}\varphi,\lambda^{2n+1}\varphi,\lambda^{2n+1}\tilde\varphi\}_{n\geq 0},\\
&&W={\Bbb C}\{ \lambda^{2n}\varphi,\lambda^{2n}\tilde\varphi\}_{n\geq 0},
\end{eqnarray*}
 where $\varphi$ and $\tilde\varphi$ are the solutions of
\[
 \left\{\frac{\sigma}{2}(\lambda\frac{d}{d\lambda}-
\frac{1}{2}(1+H))-\lambda\theta(\lambda^2)
 +\sum_{n\geq 0}\theta_{2n+1}
\ell_{2n+1}^t\left|_{{\bf
t}=0}\right.(\lambda^2)\right\}\left(\begin{array}{c}\varphi\\
\tilde\varphi\end{array}\right)=0,
\]
 having the asymptotic expansion
\[
 \left(\begin{array}{c}\varphi\\
\tilde\varphi\end{array}\right)\sim\left(\begin{array}{c}
\lambda+ \varphi_{20}+\varphi_{11}
\lambda^{-1}+\varphi_{21}\lambda^{-2}+\cdots\\
1+\tilde\varphi_{11}\lambda^{-1}+\tilde\varphi_{21}\lambda^{-2}+\cdots\end{array}
\right),\ \lambda\rightarrow\infty.
\]
Here $\theta$ and $\ell_{2n+1}$ are given by {\rm (\ref{t}), (\ref{ell})}
respectively.
\end{th}
Observe that the subspaces $V,W$ in the Sato Grassmannian are characterized by
the periodicity condition
\[
\lambda^2W\subset\lambda V\subset W
\]
an by
\[
{\cal A}V\subset\lambda W,\;{\cal A}W\subset\lambda V,
\]
where
\[
{\cal A}:=\frac{\sigma}{2}\lambda\frac{d}{d\lambda}-\lambda\theta(\lambda^2),
\]
see \cite{abs}.
Similarly, one can proof that
\begin{th}
The points in the  formal homogeneous space
\[
LSL(2,{\Bbb C})/N^+SL(2,{\Bbb C})\cong B^-SL(2,{\Bbb C})
\]
corresponding to  self--similar solutions of the potential mKdV hierarchy are
given by the solutions $\varphi,\tilde\varphi$ of
\[
 \left\{\frac{\sigma}{2}(\lambda\frac{d}{d\lambda}-
\frac{1}{2}(1+H))-\lambda\theta(\lambda^2)
+\sum_{n\geq 0}\theta_{2n+1}
\tilde\ell_{2n+1}^t\left|_{{\bf
t}=0}\right.(\lambda^2)\right\}\left(\begin{array}{c}\varphi\\
\tilde\varphi\end{array}\right)=0,
\]
 having the asymptotic expansion
\[
 \left(\begin{array}{c}\varphi\\
\tilde\varphi\end{array}\right)\sim\left(\begin{array}{c}
\lambda +\varphi_{20}+\varphi_{11}
\lambda^{-1}+\varphi_{21}\lambda^{-2}+\cdots\\
\tilde\varphi_{20}+\tilde\varphi_{11}\lambda^{-1}+\tilde\varphi_{21}\lambda^{-2}+\cdots\end{array}
\right),\ \lambda\rightarrow\infty.
\]
Here $\theta$ and $\tilde\ell_{2n+1}$ are given by {\rm (\ref{t}),
(\ref{tildell})} respectively.
\end{th}

 Given $\sigma$ one can consider $\theta(\lambda)$ as a polynomial of degree
$N$, then the functions $\varphi, \tilde\varphi$ defining the point in
$\mbox{Fl}^{(2)}$ associated to a self--similar solution of the mKdV hierarchy
depends on the parameters
\[
\{v_0,R_{n,0},\dot R_{n,0},\ddot R_{n,0}\}_{n=1}^N,
\]
where we denote by $\dot f=\partial_1 f,\, f_0=f({\bf t}=0)$  . These constants
are not independent, in
fact they fulfill the Gel'fand--Dickii relations \cite{gd}
\[
R_{n+1,0}=2\sum_{m=0}^{n-1}R_{m,0}\ddot R_{n-m,0}-\sum_{m=1}^{n-1}
\dot R_{m,0}\dot R_{n-m,0}+4u_0\sum_{m=0}^nR_{m,0}R_{n-m,0}-4\sum_{m=1}^n
R_{m,0}R_{n-m+1,0},
\]
and the string equation gives the additional constraint
\[
Av_0^2+Bv_0+C=0,
\]
where
\begin{eqnarray*}
&&A:=2\sum_{n\geq 0}\theta_{2n+1}R_{n,0},\\
&&B:=\frac{\sigma}{2}+2\sum_{n\geq 0}\theta_{2n+1}\dot R_{n,0},\\
&&C:=\sum_{n\geq 0}\theta_{2n+1}\ddot R_{n,0}+u_0A,
\end{eqnarray*}
here we have used the Miura map (\ref{miura}) connecting $u$ with $v$.
Therefore, since these are all the constraints that must be satisfied by the
constants we
conclude that our solution is parametrized by a $2N+1$--dimensional algebraic
variety
$\Sigma_{\theta}\subset{\Bbb C}^{3M+1}$. For each point in this variety
we have a subspace in the Sato periodic flag manifold $\mbox{Fl}^{(2)}$, this
map gives an inclusion
$\Sigma_{\theta}\hookrightarrow \mbox{Fl}^{(2)}$. This $2N+1$--dimensional
surface intersects the Segal--Wilson periodic flag manifold $\mbox{Fl}_0^{(2)}$
in
a discrete set, that can be labeled by $\Bbb N$, in fact   this
intersection set corresponds to an Adler--Moser rational solution to the KdV
hierarchy.

Observe that
\[
c+\frac{\sigma}{4}=Av_0+\frac{B}{2},
\]
and when $c=0$ we have the additional constraint
\[
\frac{\sigma}{4}=Av_0+\frac{B}{2}.
\]
This must be satisfied  if we are looking for self--similar solutions to the
potential mKdV hierarchy. The functions $\varphi,\tilde\varphi$ giving these
self--similar solutions   depends on the above parameters and on $w_0$, but
this  parameter is irrelevant, if $w$ is self--similar then so is any
$w+\mbox{cte}$, we can therefore fix the value of $w_0$. This analysis implies
that the moduli space is $2N$ dimensional.

 The correct number of parameters can be found directly from the string
equation. Supposing that the solution is defined at the origin we find
solutions to the string equation when $0=t_3=t_5=\cdots$, the number of
parameters needed to describe them is the dimension of the moduli. The
solutions are obtain from these initial data by applying the commuting flows on
the integrable hierarchy. Also it can be obtain by an analysis of Stokes
parameters associated to the string equation, this is the approach of \cite{m}.
Nevertheless, in our description the dimension of the moduli is obtain as the
number of parameters necesary to describe the points of the homogeneous space
associated with self--similar solutions. Therefore they have a clear
geometrical interpretation.
\section{Connection with the moduli space for the KdV hierarchy}
The  discussion in the previous section provides us  with a detailed account of
the moduli space for the string equations of UMM with border terms and also
when these border terms are absent. In this section we shall
  connect this description   with that given in \cite{gm2} for the moduli space
of self--similar solutions to the potential KdV hierarchy. The string equation
in this case is associated with the so called 2D stable quantum gravity
\cite{djm}. For  the potential KdV hierarchy the Birkhoff factorization problem
is essential. In fact the Birkhoff factorization
\[
\psi=\hat\psi_-^{-1}\cdot\hat\psi_+,
\]
where $\hat\psi_-\in L^-_1SL(2,{\Bbb C})$ and $\hat\psi_+\in L^{+}SL(2,{\Bbb
C})$, for the commuting flows $\psi({\bf t})$, can
be solved in terms of a function $p$ that parametrizes  $\hat\psi_- $ and
satisfies the potential KdV hierarchy
\[
\partial_{2n+1}p=-2R_{n+1}[u]
\]
where $u=-2\partial_1p$ is a solution of the KdV hierachy, see
\cite{sw,gm1,gm2}. This means that the projection of the commuting flows
$\psi({\bf t})$ in the Grassmannian manifold
\[
\mbox{Gr}^{(2)}\cong LSL(2,{\Bbb C})/ L^+SL(2,{\Bbb C})
\]
can be described in terms of the potential KdV hierarchy.

One can write
\[
\psi_-=\exp(aF)\cdot\hat\psi_-,
\]
where
\[
\exp(aF)=\lim_{\lambda\rightarrow\infty}\psi_-.
\]
Hence, $\psi_+=\exp(aF)\cdot\hat\psi_+$ and so \cite{gm1}
\[
\partial_1(a+p)=(a-p)^2,\; v=a-p,
\]
or
\[
a=v+p,\; u:=-2\partial_1p=\partial_1v-v^2.
\]
 The initial condition for the mKdV hierarchy can be choosen such that
\[
 g=\hat g\cdot\exp(-a_0F)\in N^-SL(2,{\Bbb C}),
\]
where $\hat g\in L^-_1SL(2,{\Bbb C})$ is the initial condition for the
corresponding solution to the potential KdV hierarchy. Thus, if
$\varphi,\tilde\varphi$ are associated with $g$, and gives the point $(V,W)$ in
$\mbox{Fl}^{(2)}$ corresponding to a   solution $v$ of the mKdV hierarchy, and
$\phi, \tilde\phi$ are associated with $ \hat g$, and thereby define a subspace
$\hat W:={\Bbb C}\{\lambda^{2n}\phi,\lambda^{2n}\tilde\phi\}_{n\geq 0}$ in
$\mbox{Gr}^{(2)}$
corresponding to a solution $u=\partial_1v-v^2$, one has
\begin{equation}
\left(\begin{array}{c}\varphi\\ \tilde\varphi\end{array}\right)=
\left(\begin{array}{cc}1&-(p_0+v_0)\\0&1\end{array}\right)
\left(\begin{array}{c}\phi\\ \tilde\phi\end{array}\right).
\label{rel}
\end{equation}
Therefore, given a solution $p$ to the potential  KdV hierarchy there is a
one--dimensional space of solutions of the mKdV hierarchy that through the
Miura map goes  to $u=-2\partial_1p$. A possible parameter for this family is
the initial value $v_0$. This has a clear geometrical interpretation \cite{w},
observe that $W=\hat W$, therefore we have the projection
\[
\begin{array}{cccc}
\pi:&\mbox{Fl}^{(2)}&\rightarrow&\mbox{Gr}^{(2)}\\
    &(V,W)          &\mapsto    &W.
\end{array}
\]
  The periodic flag manifold ${\rm Fl}^{(2)}$ is fibered over the Grassmannian
${\rm Gr}^{(2)}$, being the fiber a copy of ${\Bbb C}P^1$, \cite{ps}. The fiber
$\pi^{-1}(W)$ can be recovered from Eq. (\ref{rel}). The projection $\pi$ can
be interpreted as the Miura transformation \cite{w}, schematically this can be
encoded in the Wilson diagram
\[
\begin{CD}
\text{Fl}^{(2)} @>>> \text{mKdV}\\
@V{\pi}VV             @VV{\text{Miura map}}V\\
\text{Gr}^{(2)} @>>>   \text{KdV.}
\end{CD}
\]

For a given self--similar solution $u$ of the KdV hierarchy we have a one
parameter family of solutions to the mKdV hierarchy. The solution $u$ fixes the
Gel'fand--Dickii potentials, thus if we look for a self--similar $v$ in this
family the string equation selects two possible values  $v_{0,\pm}$. Hence,
there are only two points (associated to self--similar
solutions $v_{\pm}$) in the fiber corresponding to $u$.
Let be $c_{\pm}$ the value of $c$ for $v_{\pm}$, then
\[
c_++c_-=-\frac{\sigma}{2},
\]
and
\[
c_{\pm}=\frac{1}{2}(-\sigma\pm\Delta),
\]
where
\[
\Delta=\sqrt{B^2-4AC}
\]
is the discriminant of the equation for $v_0$.
This $\Delta$ is essentially the parameter $\Gamma$ of
the first reference of $\cite{dal3}$.

Suppose as before that $\theta$ is a polynomial of degree $N$, then as was
proof in \cite{gm2} the moduli space for the self--similar to the KdV hierarchy
is a $2N+1$--dimensional surface in
${\rm Gr}^{(2)}$, and from the above discussion we conclude that the moduli
space for the self--similar solutions of the mKdV hierarchy is double covering
of this surface, see \cite{abs}. For the mKdV case we have the following scheme
\[
\begin{CD}
{\cal M}_{\text{mKdV}}  @>>> \text{mKdV+String Eq.}\\
@V{ {\Bbb Z}_2}VV      @VV{\text{Miura map}}V\\
 {\cal M}_{\text{KdV}}  @>>>   \text{KdV+String Eq.}
\end{CD}
\]
where ${\cal M}_{\text{mKdV}}\subset \text{Fl}^{(2)}$ and ${\cal
M}_{\text{KdV}}\subset\text{Gr}^{(2)}$ denotes the moduli spaces for the
self--similar solutions of the mKdV and KdV hierarchies, respectively.

 For the potential mKdV  hierarchy, the situation is rather different.
The homogeneous space $LSL(2,{\Bbb C})/N^+SL(2,{\Bbb C})$ is a line bundle over
$\text{Fl}^{(2)}$.  This fibering is a consequence of the following fact, given
a solution $w$ of the potential mKdV hierarchy any $w+\text{constant}$ is a
solution as well. Given the initial condition $\underline{g}\in B^-SL(2,{\Bbb
C})$ one has the factorization
\[
\underline{g}=g\cdot\exp(-w_0H)
\]
where $g\in N^-SL(2,{\Bbb C})$ is the initial condition fixing the solution
$v=\partial_1 w$ of the mKdV hierarchy, and
\[
\lim_{\lambda\rightarrow\infty}\underline{g}=
\exp(-(p_0+v_0)F)\cdot\exp(-w_0 H).
\]
Given a self--similar solution $w$ any solution in the corresponding fiber is
also self--similar. Thus, we can look to the corresponding point in the base
manifold $\text{Fl}^{(2)}$, that is to $v=\partial_1 w$ self--similar solution
of the mKdV hierarchy. In this way the periodic flag manifold contains the
moduli space of self--similar solutions of the potential mKdV hierarchy.
But now we have the contraint $c=0$.   In fact, we have a subset of codimension
one in the $2N+1$--dimensional moduli space for the self--similar solutions of
the potential KdV  hierarchy
which is in a one--to--one correspondence to the self--similar solutions of the
potential mKdV hierarchy.
Hence, the moduli space is a $2N$--dimensional surface in  $\text{Fl}^{(2)}$.
Summing, when $c=0$ not only the two--folding disapears but also  not every
self--similar solution of the potential KdV hierarchy is connected to a
self--similar solution to the potential mKdV hierarchy.

 In physical terms this means that stable 2D quantum gravity \cite{dal3,djm}
(self--similar solutions of the potential KdV hierarchy) is covered twice by
the double scaling limit of the UMM with
boundary terms \cite{min}, see \cite{dal3}. But there is only a subset of
stable 2D quantum gravity corresponding to the double scaling limit of the UMM
(no border terms) \cite{per}.

Suppose that we write $\theta_{2n+3}=\hat\theta_{2n+1}$,
where we choose $\theta_1=0$. Then a possible solution to the string equation
for self--similar solutions of the mKdV hierarchy is a $v$ that satisfies
\[
\sum_{n\geq 0}\hat\theta_{2n+1}R_{n+1}+{\cal R}=0.
\]
 The corresponding $u$ is a solution to the string equation of the double
scaling limit of the HMM (translations and Galilean self--similarity in the
potential KdV hierarchy). This gives a connection between the HMM and the UMM
with border terms. Notice that $c=-\sigma/4$ and therefore the corresponding
$w$ is not self--similar. So the mentioned connection only exists when the
border terms are present, thus the HMM is not connected in this way with the
UMM.

 As an example we can analyse the case $\theta(\lambda)=-1,\,\sigma=1$. In
\cite{gm2} it was found that
\[
\phi=\lambda\left(\frac{1}{2}\frac{d\tilde\phi}{d\lambda}+\tilde\phi\right)
\]
and
\[
\tilde\phi(\lambda)\sim i
\sqrt{\frac{4\lambda}{\pi}}e^{-2\lambda}K_{\nu}(-2\lambda)\sim
\sum_{n\geq 0}(-1)^n
\frac{\Gamma(\nu+n+1/2)}{4^nn!\Gamma(\nu-n+1/2)}\lambda^{-n},
\ \ \lambda\rightarrow\infty,
\]
where
\[
\nu=\frac{\sqrt{1-16u_0}}{2},
\]
and $K_{\nu}$ is the Macdonald's function
\cite{no}.
We know that
\[
\varphi=\phi-(p_0+v_0)\tilde\phi,\; \tilde\varphi=\tilde\phi.
\]
The string equation for the potential KdV hierarchy implies
\[
p_0=-u_0,
\]
and the string equation for the mKdV hierarchy gives
\[
v_{0,\pm}=-\frac{1}{4}(-1\pm\sqrt{1-16 u_0}).
\]
 So, for a given $u$, generically we have two points $(V_{\pm},W)$ in the
periodic flag manifold, observe that when $u_0=1/16$ there is only one point
$v_0=1/4$, that is a branch point for the double covering. Now $c=v_0$.
 These solutions belongs to the Segal--Wilson periodic flag manifold  if and
only if
\begin{equation}
u_0=-\frac{m(m+1)}{4},\; m\in{\Bbb N}\cup\{0\},\label{rat}
\end{equation}
which implies
\[
v_{0,+}=-\frac{m}{2},\; v_{0,-}=\frac{m+1}{2}.
\]

When (\ref{rat}) is satisfied we are dealing with the
 rational solutions of the mKdV hierarchy, \cite{am},
$v_+=v_m$  and $v_-=-v_{m+1}$, where $v_m$ is the solution of the mKdV
hierarchy that for ${\bf t}=\{t_1,0,0,\dots\}$ is of the form
$v_m=m/(t_1-2)$. Both solutions are mapped through the Miura transformation
into the rational solution of the KdV hierarchy
that for ${\bf t}=\{t_1,0,0,\dots\}$ is of the form
$u=-m(m+1)/(t_1-2)^2$ (and $p=-m(m+1)/2(t_1-2)$). These are the well known
rational
solutions of the KdV
hierarchy, that vanish at $t_1=\infty$, analysed
by Adler and Moser \cite{am}.
For $m=0$, and  $u=0$ we have $v_+=0$ and $v_-=-1/(t_1-2)$; for $m=1$, and
$u=-2/(t_1-2)^2$ one has $v_+=1/(t_1-2)$ and
$v_-=(t_3-2(t_1-2)^3)/((t_1-2)((t_1-2)^3+t_3))$, observe that $u$ only depends
on $t_1$ and that $v_-$ depends also upon $t_3$.

 For an arbitrary $u_0$ we have two points in the Sato periodic flag manifold,
so there is a one--di\-men\-sional complex curve in this space giving scaling
self--similar solutions.

Observe that if (\ref{rat}) is satisfied then $\nu=m+1/2$, and $\tilde\varphi$
is the following polynomial in $\lambda^{-1}$
\[
\tilde\varphi(\lambda)=\lambda^{m+1}e^{-2\lambda}
\left(\frac{1}{2\lambda}\frac{d}{d\lambda}\right)^{m+1}e^{2\lambda}.
\]

If we look for self--similar solutions of the potential mKdV hierarchy we need
$c=0$, hence $v_0=0$ and $u_0=0$, which gives $v=0$ and $w={\rm cte}$. In this
case the solution is unique and trivial.
\section{Acknowledgements}
One of us MM like to thank Prof. Francisco Guil for initial collaboration and
PG like to thank Prof. Pierre van Moerbeke for enlighting discussion. We are
also grateful to Dr. Sacha Sardo Infirri for stimulating discussion.

\end{document}